\documentclass[12pt,english,twoside]{article}
\RequirePackage{amsmath,amssymb,amsfonts,textcomp,yhmath}
\RequirePackage{times} 
\RequirePackage{cases} 
\RequirePackage[pdftex]{hyperref}\hypersetup{backref,implicit=true,bookmarks=true,
colorlinks=true, linkcolor=blue,filecolor=blue,
pagecolor=blue,urlcolor=blue,pdfpagemode=UseOutlines}

\newtheorem{theorem}{Theorem}[section]

\newtheorem{prev}{Proof}[section]

\begin{document}
\markboth{\small{A. Nangue}}{\tiny{Global existence of solutions to
the Einstein-massive scalar field equations with a cosmological
constant on the flat Robertson-Walker space-time}}
 \centerline{}


\begin{center}
\Large{{\bf Global existence of solutions to the Einstein-massive
scalar field equations with a cosmological constant for a perfect
fluid on the flat Robertson-Walker space-time}}
\end{center}

\centerline{}
 \centerline{\bf {Alexis Nangue}}
 \centerline{University of Maroua}
\centerline{Higher Teacher's Training College,}
  \centerline{Department of Mathematics, PO.Box 55, Maroua, Cameroon}
 \centerline{alexnanga02@yahoo.fr}

\begin{abstract}
In many cases a massive nonlinear scalar field can lead to
accelerated expansion in cosmological models. This paper contains
mathematical results on this subject for flat Robertson-Walker
space-time. Global existence to the coupled Einstein-massive scalar
field system which rules the dynamics of a kind of pure matter in
the presence of a massive scalar field and cosmological constant is
proved, under the assumption that the scalar field $\phi$ is a
non-decreasing function, in Robertson-Walker\footnote{\tiny{For
convenience instead of Friedmann-Lemaître-Robertson-Walker
space-time we just write Robertson-Walker space-time.}} space-time;
asymptotic behaviour are investigated in the case of a cosmological
constant bounded from below by a strictly negative constant
depending only on the massive scalar field.
\end{abstract}
\noindent
{\bf MR Subject Classification}: 83C05, 83F05 \\


\section{Introduction}
Scalar fields currently play a remarkable role in the construction
of cosmological scenarios aiming to describe the structure and
evolution of the early universe. Cosmological models with
accelerated expansion are presently of great astrophysical interest.
Two main themes are inflation, which concerns the very early
universe, and the accelerated cosmological expansion at the present
epoch as evidenced, for instance, by supernova observations.
Generally the simplest way of obtaining accelerated expansion within
General Relativity is a positive cosmological constant. A more
sophisticated variant is the presence of a nonlinear scalar field.
In this paper we consider both cosmological constant and scalar
field and this leads to interesting new results. For the two themes
mentioned above this scalar field is known as the inflaton and
quintessence, respectively. A cosmological constant alone may not be
an alternative for describing the real universe if the observational
data indicate different amounts of acceleration at different times.
Evidence of this kind based on supernovae-Ia data is presented
in\cite{Alam}. There is at present no good physical understanding of
the exact nature of the inflaton or quintessence and it may even be
that both are the same field, producing different phenomena at
different times.

 Global dynamics of various matter fields coupled to
the Einstein equations remain an active research area in General
Relativity. The Robertson-Walker space-time is considered to be the
basic space-time in cosmology, where homogeneous phenomena such as
the one we consider in the present paper are relevant. There are
several reasons why it is of interest to consider the Einstein
equations with cosmological constant and to couple the equations to
a massive nonlinear scalar field.
\begin{enumerate}
    \item[$\bullet$]Astrophysical observations have made evident the
    fact that, even in the presence of material bodies, the
    gravitational field can propagate through space at the speed of
    light, analogously to electromagnetic waves. A mathematical way
    to model this phenomenon is to couple a scalar field to the
    Einstein equations. Let us recall that the Nobel Prize of
    physics 1993 was awarded for works on this subject. More details
    on this question can be found in \cite{LDE}, \cite{AD2}. In fact,
     several authors realized the interest of coupling scalar field
     to other fields equations; see for instance \cite{Chris},
     \cite{HL1}, \cite{AD1}, \cite{AD2}, \cite{Tegang}, \cite{AN1}.
    \item[$\bullet$]Now our motivation for considering the Einstein equations
with cosmological constant $\Lambda$ is due to the fact that
astrophysical observations, based on luminosity via redshift plots
of some far away objets such as Supernovae-Ia \cite{Alam}, have made
evident the fact that the expansion of the universe is accelerating.
A classical mathematical tool to model this phenomenon is to include
the cosmological constant $\Lambda$ in the Einstein Equations.
 In recent years, cosmological models with
    accelerated expansion have become a particularly active
    rechearch topic; see for instance \cite{HL2}, \cite{NN1},
    \cite{NN2}, \cite{NN3}, \cite{NN4}, \cite{NN5}.
\end{enumerate}
Also recall that the recent Nobel prize of Physics, 2011, was
awarded to three Astrophysicists for their advanced research on this
phenomenon of accelerated expansion of the universe.

In fact, we must point out that, the notion of "dark energy" was
introduced in order to provide a physical explanation to universe
expansion phenomenon, but the physical structure of this
hypothetical form of energy which is unknown in the laboratories
remains an open question in modern cosmology; so is the question of
"dark matter". Also notice that the scalar fields are considered to
be a mechanism producing accelerated models, non only in
"inflation", which is a variant of the Big-Bang theory including now
a very short period of very high acceleration, but also in the
primordial universe. Formally the Robertson-Walker model with scalar
field is obtained if we add to the matter content of the classical
Friedmann universe a perfect fluid with energy density
$\epsilon:=\frac{1}{2}\dot{\phi}^{2}+\frac{1}{2}m^2\phi^2$ and
pressure $q:=\frac{1}{2}\dot{\phi}^{2}-\frac{1}{2}m^2\phi^2$.
However, this fluid violates the strong energy condition,i.e.,
$\epsilon+3q=2\dot{\phi}^{2}-m^2\phi^2$ may be negative. It is
precisely this violation that leads to inflation in the very early
universe \cite{Kolb}.

In General Relativity the evolution of Robertson-Walker models of
ordinary matter with a massive scalar  are governed, Following
\cite{hawking}, by equations:

\begin{equation}\label{2}
R_{\alpha\beta}-\frac{1}{2}R g_{\alpha\beta}+\Lambda
g_{\alpha\beta}= 8\pi (T_{\alpha\beta}+\tau_{\alpha\beta})
\end{equation}
\begin{equation}\label{3}
T_{\alpha\beta}= (\rho + p)u_{\alpha}u_{\alpha} + pg_{\alpha\beta}
\end{equation}
\begin{equation}\label{4}
\tau_{\alpha\beta}= \nabla_{\alpha}\phi \nabla_{\alpha}\phi
-\frac{1}{2}(\nabla^{\lambda}\phi \nabla_{\lambda}\phi + m^2\phi^2),
\end{equation}
what we call in the present paper  the Einstein-massive scalar field
system; where:

\begin{itemize}
    \item[-] (\ref{2}) are the Einstein equations\footnote
    {\tiny{Units have been chosen so that $c=1=G$.}} for the metric tensor
    $g=(g_{\alpha\beta})$, which represents the gravitational field.
    $g$ in this case depends on a unknown single
    positive real-valued function $a=a(t)$, called the cosmological
    expansion factor. $R_{\alpha\beta}$ is the Ricci tensor,
    contracted of the curvature tensor.
    $R=g^{\alpha\beta}R_{\alpha\beta}$ is the scalar curvature,
    contracted of the Ricci tensor.
    \item[-]The ordinary matter is modeling by (\ref{3}),  which represents the general expression of the
    stress-matter-energy
    tensor of a relativistic perfect fluid in the choosen signature
    of $g$ ; in which $\rho\geq0$ and $p\geq 0$ are unknown functions
    of the single variable $t$, representing respectively the matter density and
    pressure. We consider a perfect fluid of pure radiation type
    which means that $p=(1-\gamma)\rho$, with $0\leq \gamma \leq 2$
    \footnote{\tiny{In this work we consider the case where $\gamma=\frac{4}{3}$.}}.
    In order to simplify, we consider a co-moving fluid : this implies
    that $u_{i}=u^{i}=0$ where $u=(u_{\alpha})$ is a time-like unit
    vector represents the velocity.
    \item[-] (\ref{4}) represents the stress-matter-energy tensor
    associated to a non-decreasing
    \footnote{\tiny{That is : $\dot{\phi}\geq 0$ where overdot denotes differentiation
   with respect to time $t$.}}
     massive scalar field $\phi$, which is as $\rho$ a real-valued
     function of $t$ and finally $\nabla_{\alpha}$ is the covariant
     derivative in $g$.
\end{itemize}

Now, recall that, solving the Einstein equations is determining both
the gravitational field and its sources : this means that we have to
determine every unknown function introduced above, namely: $a$,
$\rho$ and $\phi$. Notice that the spatially homogeneous coupled
Einstein-massive scalar field system turns out to be a non-linear
second differential system in $a$, $\phi$ and $\rho$. Also notice
that, what we call global solution in the present paper, is a
solution defined all over
the interval $[0, +\infty[$.\\
In this paper, we prove that if the initial value of the Hubble
variable $u$ is strictly negative and if $\Lambda > - \alpha^{2}$,
where $\alpha > 0$ is a constant depending only on the
potential\footnote{\tiny{The potential of the sacalar field is
defined here by $V(\phi)=\frac{1}{2}m^2\phi^2$.}} of the scalar
field, then the coupled Einstein-massive scalar field system, with
cosmological constant, has a global in time-solution. This result
extends and completes those of \cite{AN1}. We investigate the
asymptotic behaviour which reveals an exponential growth of the
gravitational potentials, confirming the accelerated expansion of
the univers.\\
The paper is organized as follows:
\begin{enumerate}
    \item[$\circ$] In section 2, we write the Einstein-massive scalar field system in explicit terms.
    \item[$\circ$] In section 3, we introduce the Cauchy problem and
    we prove the local and global existence of solutions.
    \item[$\circ$]In section 4, we study the asymptotic behaviour.
\end{enumerate}

\section{Einstein-massive scalar field equations in $a$, $\rho$, $\phi$}
In this section we are going to write the equations (\ref{2}) in
explicit terms, next we proceed by a change of unknown functions.
The evolution of Robertson-Walker models with massive scalar field
$\phi$, cosmological constant $\Lambda$  and ordinary matter, which
described by a perfect fluid with matter density $\rho$, are
governed following \cite{AN3}, \cite{AN1} and  \cite{AN2}, by the
constraint equation named Hamiltonian equation,

\begin{equation}\label{12}
3\Big(\frac{\dot{a}}{a}\Big)^{2}-\Lambda=4\pi(\dot{\phi}^2+m^2\phi^2+2\rho),
\end{equation}
the evolution equation,
\begin{equation}\label{13}
 2\Big(\frac{\ddot{a}}{a}\Big)+\Big(\frac{\dot{a}}{a}\Big)^{2}-\Lambda=-4\pi(\dot{\phi}^{2}-m^2\phi^2+\frac{2\rho}{3}),
\end{equation}
the equations in $\phi$ and $\rho$ resulting from conservation
equation,
\begin{equation}\label{14}
  \dot{\phi}\ddot{\phi}+3\frac{\dot{a}}{a}\dot{\phi}^{2}+m^2\phi\dot{\phi}=0,
\end{equation}
\begin{equation}\label{15}
    \dot{\rho}+4\frac{\dot{a}}{a}\rho=0
\end{equation}
In next paragraphs, we study the global existence of solutions $a$,
$\rho$ and $\phi$ to the coupled system (\ref{13}),  (\ref{14}),
(\ref{15}) under constraint (\ref{12}). For this aim, we make a
change of unknown functions in order to deduce an equivalent first
order differential system to which standard theory applies. We set:
\begin{equation}\label{30}
    u=\frac{\dot{a}}{a} \; ; \; v=\frac{1}{a^2} \; ; \;
    \psi=\frac{1}{2}\dot{\phi}^2.
\end{equation}
$u$ is named the Hubble variable; we deduce from (\ref{30}):
\begin{equation}\label{30'}
\frac{\ddot{a}}{a}=\dot{u}+u^2 \; ; \; \dot{v}=-2uv.
\end{equation}
We choose to look for a $\mathcal{C}^{2}$-non-decreasing massive
scalar field, $\dot{\phi}\geq 0$, (\ref{30}) then gives:
\begin{equation}\label{31}
   \dot{\phi}=\sqrt{2}\psi^{\frac{1}{2}}.
\end{equation}
According to (\ref{12}) and (\ref{30})we  deduce from (\ref{13}),
(\ref{14}), (\ref{15}) (\ref{30}) and (\ref{30'}) the equivalent
first order differential system:

\begin{numcases}
\strut
\dot{u}= -\frac{3}{2}u^2+\frac{\Lambda}{2}-4\pi(\psi-\frac{1}{2}m^2\phi^2+\frac{1}{3}\rho)\label{37}\\
\dot{v}= -2uv            \label{38}\\
\dot{\psi}=-6u\psi-\sqrt{2}m^2\Phi\psi^{\frac{1}{2}} \label{39}\\
\dot{\phi}=\sqrt{2} \psi^{\frac{1}{2}}         \label{40}\\
\dot{\rho}= -4u \rho,  \label{40'}
\end{numcases}

to study, subject to the constraint:
\begin{equation}\label{41}
    3u^2-\Lambda=8\pi(\psi+\frac{1}{2}m^2\phi^2+\rho)
\end{equation}
\section{Global existence of solutions}
\subsection{Cauchy problem and constraints}
Let $a_{0} > 0$, $\rho_{0}\geq 0$, $b_{0}$, $\dot{\phi}_{0}>0$,
$\phi_{0} \in \mathbb{R}$ be given real numbers. We look for
solutions $a$, $\rho$ and $\phi$ of the Einstein-massive scalar
field system over $[0, T[$, $T\leq +\infty $, satisfying:
\begin{equation}\label{34}
a(0)=a_{0}\; ; \;\dot{a}(0)=b_{0}\; ; \;\phi(0)=\phi_{0}\; ;
\;\dot{\phi}(0)=\dot{\phi}_{0}\; ; \;\rho(0)=\rho_{0}.
\end{equation}
Our objective now is to prove the existence of global solutions $a$,
$\rho$ and $\phi$ defined over the whole interval $[0, +\infty[$,
and satisfying (\ref{34}) called initial conditions, the given
numbers $a_{0}$, $b_{0}$, $\phi_{0}$, $\dot{\phi}_{0}$, $\rho_{0}$
being the initial data.\\
It is well know that equation (\ref{12}) called the Hamiltonian
constraint is satisfied all over the domain of the solutions of
equation (\ref{13}) called evolution equation, if and only if
equation (\ref{12}) is satisfied at $t=0$ i.e. given  (\ref{34}) if
the initial data satisfy:
\begin{equation}\label{35}
3\Big(\frac{b_{0}}{a_{0}}\Big)^{2}-\Lambda=8\pi(\dot{\phi}_{0}^2+m^2\phi_{0}^2+2\rho_{0}),
\end{equation}
called the initial constraint. Notice that if $a_{0}$, $b_{0}$,
$\phi_{0}$, $\rho_{0}$ , $\dot{\phi}_{0}$ and $\Lambda$ are given
such that $8\pi(\dot{\phi}_{0}^2+m^2\phi_{0}^2+2\rho_{0})+\Lambda
\geq 0$, then (\ref{35}) gives two possible choices of $b_{0}$,
namely $b_{0}\geq 0$ and $b_{0}< 0$. As we will see, this choice of
the sign of $b_{0}=\dot{a}_{0}$ called the initial velocity of
expansion, will play a key role as far as the global existence of
solutions is concerned. In what follows, we suppose that the initial
constraint (\ref{35}) holds. Now we are going to study the
equivalent first order differential system (\ref{37})-(\ref{40'}),
under the constraint (\ref{41}) and with the initial conditions at
$t=0$, provided by (\ref{34}) and (\ref{30}):
\begin{equation}\label{36}
u(0):=u_{0}=\frac{b_0}{a_0}\; ;
\;\dot{v}(0):=v_{0}=\frac{1}{a^{2}_{0}}\; ; \;\phi(0)=\phi_{0}\; ;
\;\psi(0):=\psi_{0}=\dot{\phi}_{0}\; ; \;\rho(0)=\rho_{0}.
\end{equation}
\subsection{Local and global existence of solutions}
Notice that, by standard theory on the first order differential
systems, the Cauchy problem for the (autonomous) system
(\ref{37})-(\ref{40'}) always admits a unique local solution. What
we want to know now is wether or not, this solution is global. Now,
also following the standard theory on the first order differential
systems, to show that the solution is global, it will be enough if
we prove that any solution of the cauchy problem remains uniformly
bounded.
\begin{theorem}\label{theo1}\hspace*{2cm}\\
Let $\Lambda > - 4\pi m^2\phi_{0}^{2}$ be given and suppose
$\phi_{0} >0$ and $u_{0} > 0$. Then the initial value problem for
the Einstein-massive scalar field system (\ref{37})-(\ref{40'}) has
a unique global solution defined all over the interval $[0.
+\infty[$.
\end{theorem}

\begin{prev}\hspace*{2cm}\\
It will be enough if we could prove, given the evolution system
(\ref{37})-(\ref{40'}) that, if each of the functions : $u$, $v$,
$\psi$, $\phi$ and $\rho$ is uniformly bounded over every bounded
interval.(for instance $[0, T^{*}[$, where  $T^{*}< +\infty$)
\begin{enumerate}
    \item[$\bullet$] Case of $u$.\\
Using the evolution equation (\ref{37}) in $u$, we obtain:
\begin{equation}\label{42}
   \dot{u}=-\frac{3}{2}u^{2}+\frac{\Lambda}{2}+2\pi
   m^2\phi^2-4\pi\psi-\frac{4}{3}\pi\rho.
\end{equation}

But since $\psi \geq 0$, $\rho \geq 0$ (\ref{42}) gives:
\begin{equation}\label{43}
   \dot{u} \leq -\frac{3}{2}u^{2}+\frac{\Lambda}{2}+2\pi
   m^2\phi^2.
\end{equation}
Considering now the Hamiltonian constraint (\ref{41}) and since
$\psi \geq 0$, $\rho \geq 0$ we have:
\begin{equation}\label{44}
-3u^{2}+\Lambda+4\pi
   m^2\phi^2 \leq 0,
\end{equation}

(\ref{43}) then implies:
\begin{equation}\label{45}
    \dot{u} \leq 0
\end{equation}
so that $u$ is decreasing. We also deduce from (\ref{44}) since by
assumption we have $\dot{\phi}\geq 0 $, then $\phi \geq \phi(0)> 0$,
that:
\begin{equation}\label{46}
   u^{2}\geq \frac{\Lambda}{3}+\frac{4}{3}\pi m^2\phi^{2}_{0}.
\end{equation}

But by hypothesis the r.h.s of (\ref{46}) is strictly positive. So,
since $u$ is continuous, by the mean value theorem, (\ref{46})
implies:

\begin{equation}\label{47'}
u \leq - (\frac{\Lambda}{3}+\frac{4}{3}\pi m^2\phi^{2}_{0})^{\frac{1}{2}}
\end{equation}
or
\begin{equation}\label{47}
u \geq (\frac{\Lambda}{3}+\frac{4}{3}\pi m^2\phi^{2}_{0})^{\frac{1}{2}} > 0.
\end{equation}

Also by hypothesis, $u(0)> 0$, then only (\ref{47}) holds : moreover
(\ref{45}) implies $u\leq u(0)$ and we have inequalities:
\begin{equation}\label{48}
(\frac{\Lambda}{3}+\frac{4}{3}\pi m^2\phi^{2}_{0})^{\frac{1}{2}} \leq u \leq
u(0),
\end{equation}
which show that $u$ is uniformly bounded.
    \item[$\bullet$]Case of $v$.\\
Now by (\ref{38}), and given (\ref{48}) and $v>0$, we have $\dot{v}
< 0$ so that $v$ is a decreasing and positive function, hence:
\begin{equation}\label{49}
    0<v<\frac{1}{a^{2}_{0}}.
\end{equation}
Therefore $v$ is also uniformly bounded.
    \item[$\bullet$]Case of $\phi$.\\
    We deduce at once from (\ref{44}) and using  (\ref{48}) that
    $\phi$ is bounded.
    \item[$\bullet$]Case of $\psi$.\\
    Let us consider once more the Hamiltonian constraint on the
    following form:
    \begin{equation}\label{50}
        3u^{2}-\Lambda-4\pi m^2\phi^2-8\pi\rho=8\pi\psi ;
    \end{equation}
since the l.h.s of (\ref{50}) is bounded then $\psi$ is also
bounded.
    \item[$\bullet$]Case of $\rho$.\\
(\ref{40'}) is a linear first order 0.d.e in $\rho$, which solves at
once over $[0, t]$, $ t> 0$ to gives:
\begin{equation}\label{50'}
    \rho(t)=\rho(0)\exp(-4\int^{t}_{0}u(s)ds).
\end{equation}
(\ref{50'}) shows that, since by (\ref{48}), $u>0$, that
$|\rho(t)|\leq|\rho_{0}|$.\\
This completes the proof of Theorem \ref{theo1}.
\end{enumerate}
\end{prev}
\section{Asymptotic behaviour}
The mean curvature H of the space-time is defined by
$H=-g^{ij}k_{ij}$, where $k_{ij}$, is the fundamental form of the
hypersurfaces of constant time defined in the present case by
$k_{ij}=-\frac{1}{2}\partial_{t}g_{ij}$; Hence in our paper:
\begin{equation}\label{50'}
    H=-g^{ij}k_{ij}=3\frac{\dot{a}}{a}=3u.
\end{equation}
Now we consider the global solution over $[0, +\infty[$ and we
investigate the asymptotic behaviour of the different elements at
late time. We introduce as in \cite{AN1} the following quantity
which plays a key role:
\begin{equation}\label{51}
    Q=H^2-24\pi T_{00}-3\Lambda.
\end{equation}
Notice that expression (\ref{51}) of Q then shows, using Hamiltonian
constraint (\ref{41}) that $Q \geq 0$. Let us point out first of all
that, the quantity Q plays a key role in General Relativity, and in
the presence of the massive scalar field, it stands for the
quantities S in \cite{HL1}, Z in \cite{KIT1}, $\tilde{S}$ in
\cite{Moss}, $\overline{S}$ in \cite{AD1} and reduces to $H^2 \pm
3\Lambda $ in \cite{Wald} which deals with the
case of zero scalar field.\\
At late time we have the following asymptotic behaviour:
\begin{theorem}\label{theo2}\hspace*{2cm}\\
On the assumptions of theorem \ref{theo1}:
\begin{eqnarray}
   Q&=& \mathcal{O}(e^{-3\nu t}) \label{52} \\
  \rho &=&  \mathcal{O}(e^{-3\nu t})\label{53}\\
   \dot{\phi}^{2}&=&  \mathcal{O}(e^{-3\nu t})\label{54}\\
\phi^{2}   &\longrightarrow& L > 0 \label{55}\\
 T_{00}  &\longrightarrow& \frac{m^2}{2}L \label{56}\\
 H &\longrightarrow & (3C_{0})^{\frac{1}{2}} \label{57}\\
  a &\longrightarrow& +\infty .\label{58}
\end{eqnarray}
Where:
\begin{equation*}
\nu= [ \frac{1}{3}(\Lambda + 4\pi m^2 \phi_{0}^{2})]^{\frac{1}{2}}
\; ; \; C_{0}=\Lambda + 4 \pi m^2 L .
\end{equation*}
\end{theorem}
\begin{prev}\hspace*{2cm}\\
\begin{enumerate}
    \item[$\bullet$]We have, using the expression
\begin{equation}\label{59}
T_{00}=\psi +\frac{1}{2}m^2\phi^2,
\end{equation}
then, the evolution equations (\ref{39}) and (\ref{40}) in $\psi$
and $\phi$ give:
\begin{equation}\label{60}
    \dot{T}_{00}=-2H \psi.
\end{equation}
Expression (\ref{51}) of Q then gives, using (\ref{60})
\begin{equation}\label{61}
    \dot{Q}=2H(\dot{H}+24\pi\psi)
\end{equation}
then using (\ref{37}) and (\ref{50'}) we obtain:
\begin{equation}\label{62}
    \dot{Q}=-H(H^{2}-3\Lambda-24\pi T_{00}+ 8\pi \rho ).
\end{equation}
But since $8\pi \rho \geq 0$ and $-H < 0 $ and given the expression
of Q, (\ref{62}) leads to:
\begin{eqnarray}
 \dot{Q}  &\leq& -H Q.\label{63}
\end{eqnarray}
Integrating (\ref{63}) over $[0, t]$, $t>0$ yields:
\begin{equation*}
    0 \leq Q \leq Q_{0}\exp(\int_{0}^{t}-H ds);
\end{equation*}
and (\ref{48}) gives:
\begin{equation*}
0 \leq Q \leq Q_{0}\exp(-3t(\frac{1}{3}(\Lambda + 4\pi
m^2\phi^{2}_{0}))^{\frac{1}{2}})
\end{equation*}
and (\ref{52}) follows.
    \item[$\bullet$]Using the Hamiltonian constraint, the results
    (\ref{53}) and (\ref{54}) (since $\dot{\phi}^{2}=2\psi$) are
    direct consequences of (\ref{52}).
    \item[$\bullet$]The evolution equations in $\phi$ and $\psi$
    give:
    \begin{equation}\label{64}
 m^2\phi\dot{\phi}+\dot{\psi}=\frac{d}{dt}[\psi+\frac{1}{2}m^2\phi^2]=-2H\psi.
    \end{equation}
But since $H> 0$ and $\psi >0 $, (\ref{64}) implies that the
quantity $\psi + \frac{1}{2}m^2 \phi^2 $ is an decreasing function.
We then deduce that:
\begin{equation}\label{65}
\frac{m^2}{2}\phi^2\leq \frac{m^2}{2}\phi^2+ \psi \leq
\frac{m^2}{2}\phi^{2}_{0}+\psi_{0}.
\end{equation}
Hence $\phi^2$ is bounded. But the evolution equation in $\phi$
shows that $\dot{\phi}> 0$; then $\phi \geq \phi_{0} > 0$ and since
$\frac{d}{dt}(\phi^2)=2\phi\dot{\phi} > 0$, $\phi^2$ is an
increasing function. $\phi^2$ being positive, increasing and bounded
has a strictly positive limit, i.e. there exists $L> 0$ such that:
\begin{equation*}
\phi^{2} \longrightarrow L
\end{equation*}
with:
\begin{equation*}
    \phi^{2} \leq L,
\end{equation*}
hence (\ref{55}) follows.
    \item[$\bullet$] (\ref{56}) follows from (\ref{54}), (\ref{55}) and
    (\ref{59}).
    \item[$\bullet$]To prove (\ref{57}) since by (\ref{52}) $Q\longrightarrow
    0$, its expression (\ref{51}) shows, using (\ref{52}) that:
    \begin{equation}\label{67}
    H^2-3\Lambda \longrightarrow 12\pi m^2 L.
    \end{equation}
Hence:
\begin{equation*}
H^{2}-(3\Lambda+12\pi m^2L)=[H-(3\Lambda+12\pi m^2L
)^{\frac{1}{2}}][H+(3\Lambda+12\pi m^2L)^{\frac{1}{2}}]
\longrightarrow 0.
\end{equation*}
But by (\ref{48}), $H> 0$; so :
\begin{equation*}
H+(3\Lambda+12\pi m^2L)^{\frac{1}{2}}>(3\Lambda+12\pi m^2L)>0,
\end{equation*}
we then deduce that:
\begin{equation}\label{68'}
    H \longrightarrow (3C_{0})^{\frac{1}{2}}.
\end{equation}
\item[$\bullet$] We deduce directly from (\ref{41}) and (\ref{47})
since $\dot{a}> 0$ that : $u=\dfrac{\dot{a}}{a}\geq C_{1}> 0$, where
$C_{1}$ is a constant. Then integrating over $[0, t]$, $t>0$, we
have:
\begin{equation}\label{68}
a(t)\geq a_{0} \exp(C_{0}t).
\end{equation}
\end{enumerate}
This completes the proof of Theorem \ref{theo2}
\end{prev}
\section*{Concluding remarks}
\begin{enumerate}
    \item[$\star$] It follows from (\ref{68}) that $a(t)\longrightarrow +\infty$ as $t \longrightarrow +\infty$
    which shows an exponential growth of the cosmological expansion
    factor and (\ref{68}) also confirms mathematically the
    acceleration phenomenon of the expansion of the universe.
    \item[$\star$]Theorem \ref{theo1} extends strictly the results
    of \cite{AN1} which proved global existence in the case $\Lambda >\Lambda_{0}>
    0$. This new interesting result is due to the presence of the
    massive scalar field. In work in progress we aim to study the
    cases of positively and negatively curved Robertson-Walker
    spacetimes and geodesic completness.
\end{enumerate}
\section*{Acknowledgements}
I acknowledge with thanks the support of Higher Teacher's Training
College of University of Maroua where this paper was initiated,
prepared and finalized.


\begin{thebibliography}{122}
\bibitem{Alam}
U. Alam and al, \emph{Is there supernova evidence for dark energy
metamorphosis?}, (1983). Preprint: astro-ph/0311364.
\bibitem{Chris} D. Christodoulou, \emph{Bounded variation solution of the
spherically symmetric Einstein-scalar-fields equations},\\
Comm. Pure. Appl. Math 46 (1993) 1131-1220.
\bibitem{LDE} L. Derone, \emph{Le syst\`eme de d\'etection de
l'exp\'erience VIRGO d\'edi\'ee \'a la recherche d'ondes
gravitationnelles}.\\ Th\`ese
(1999):http://fr.wikipedia.org/wiki/portail.ondes gravitationnelles.
\bibitem{hawking}
S.w. Hawking and F.R. Ellis, \emph{The large scale of space-time},
(Cambridge Monographs and Maths) Cambridge:Cambridge University
Press, (1973).
\bibitem{KIT1}
Kitada, Y. and Maeda, K. \emph{Cosmic no-hair theorem in homogeneous
spacetimes I Bianchi models}. Class. Quantum Grav. 10, 703-734
(1993).
\bibitem{Kolb}
E.W.Kolb and M.S.Turner, \emph{The Early Universe}, (Addison Wesley,
1990); A. Linde, Particle Physics and Inflationary Cosmology,
(Harwood, 1990).
\bibitem{HL2}
H. Lee, \emph{Asymptotic behaviour of the Einstein -vlasov system
with a positive  cosmological constant},
 Math. Proc. Comb. Phil. Soc.137, 495-509 (2004).
\bibitem{HL1}
H. Lee, \emph{The Einstein-Vlasov system with a scalar field}: Ann.
H. Poincar\'e 6, 687-723 (2005).
\bibitem{Moss}
I. Moss and V.Sahni,  \emph{Anisotropy in the chaotic inflationary
universe}, Phys. Lett. B \textbf{178}, 159-162, (1983).
\bibitem{AN3}
N. Noutchegueme and A. Nangue \emph{Einstein-Maxwell-Massive Scalar
Field System in 3+1 formulation on Bianchi Spacetimes type I-VIII}.
Preprint:http://arxiv.org/abs/gr-qc/1205.0469V1 (2012).
\bibitem{AN1}
N. Noutchegueme and A. Nangue, \emph{Global Existence to the
Einstein-Scalar Field system on the Robertson-Walker space-times
with hyperbolic and spherical symmetries}, J.Hyperbolic Differential
Equation 7(1), 69-83, (2010).
\bibitem{AN2}
N. Noutchegueme and A. Nangue \emph{Global dynamics for a
relativistic charged matter in the presence of a massive scalar
field and the presence of a cosmological constant on Bianchi
spacetimes}, J. Math. Phys. 53, 102502, (2012).
\bibitem{NN1}
N.Noutchegueme and E. Takou, \emph{Global existence of solutions for
the Einstein-Boltzmann system  with cosmological constant in a
Friedman-Robertson-Walker space-time},  Comm. Math. Sci 4(2)
,295-314, (2006).
\bibitem{NN3}
N.Noutchegueme and G. Chendjou, \emph{Global solutions to the
Einstein equations with cosmological constant on
Friedman-Robertson-Walker space-time with plane, hyperbolic and
spherical symmetries}, Comm. Math. Sci 6(3),595-610, (2008).
\bibitem{Tets}
N. Noutchegueme and E. M. Tetsadjio. \emph{Global dynamics for a
collisionless charged plasma in Bianchi spacetimes}. Class. Quantum
Grav 26 195001, (2009).
\bibitem{NN2}
N.Noutchegueme and al, \emph{Global existence of solutions to the
Einstein equations with cosmological constant for a perfect
relativistic fluid on a Bianchi type I space-time}, Comm. Math. Sci
6(2) ,695-705, (2008).
\bibitem{AD2}
A.D. Rendall: \emph{on the nature of singularities in plane symmetry
scalar field cosmologies}, Gen. Relativity and gravitation 27,
213-221, (1995).
\bibitem{AD1}
A.D. Rendall: \emph{Accelerated cosmological expansion due to a
scalar field whose potential has a positive lower bound}. Class.
Quantum Grav. 21, 2445-2454 (2004).
\bibitem{NN4}
N.Straumann, \emph{On the cosmological constant problems and the
astronomical evidence for the homogeneous energy density with
negative pressure in, Vacuum Energy}, Renormalisation eds. B.
Duplantier and V. Rivasseau. (Birkhausser, Basel, (2003).
\bibitem{NN5}
S.B.Tchapnda and N.Noutchegueme: \emph{the surface symmetric
Einstein-Vlasov system with cosmological constant},
Math.Proc.Cambridge Phil.Soc 138, 541-724, (2005).
\bibitem{Tegang}
D.Tegankong, N.Noutchegueme and  A.D.Rendall, \emph{Local existence
and continuation criteria for solutions of the
Einstein-Vlasov-Scalar field system}, J.Hyperbolic Differential
Equations 1(4) 691-724, (2004).
\bibitem{Wald}
Wald, R, \emph{Asymptotic behaviour of homogeneous cosmological
models in the presence of a positive cosmological constant}.
Phys.Review. D 28, 2118-2120, (1983).
\bibitem{Wald2}
Wald, R. 1984 \emph{General Relativity} (Chicago II: University of
Chicago Press).
\end{thebibliography}
\end{document}